# Mid-infrared dual-comb spectroscopy with low drive-power on-chip sources

Lukasz A. Sterczewski[1,2,3†], Jonas Westberg[1,†], Mahmood Bagheri[2], Clifford Frez[2], Igor Vurgaftman[4], Chadwick L. Canedy[4], William W. Bewley[4], Charles D. Merritt[4], Chul Soo Kim[4], Mijin Kim[5], Jerry R. Meyer[4], and Gerard Wysocki[1,*]

[1]Department of Electrical Engineering, Princeton University, Princeton, New Jersey 08544, USA
[2]Jet Propulsion Laboratory, California Institute of Technology, Pasadena, CA 91109, USA
[3]Faculty of Electronics, Wroclaw University of Science and Technology, Wroclaw 50370, Poland
[4]Naval Research Laboratory, Code 5613, Washington, DC 20375, USA
[5]Sotera Defense Solutions, Inc., 7230 Lee DeForest Drive, Suite 100, Columbia MD 21046, USA

[†]These authors contributed equally to this work.
[*]Corresponding author: gwysocki@princeton.edu

**Two semiconductor optical frequency combs consuming less than 1 W of electrical power are used to demonstrate high-sensitivity mid-infrared dual-comb spectroscopy in the important 3-4 μm spectral region. The devices are 4 millimeters long by 4 microns wide, and each emits 8 mW of average optical power. The spectroscopic sensing performance is demonstrated by measurements of methane and hydrogen chloride with a spectral coverage of 33 cm$^{-1}$ (1 THz), 0.32 cm$^{-1}$ (9.7 GHz) frequency sampling interval, and peak signal-to-noise ratio of ~100 at 100 μs integration time. The monolithic design, low drive power, and direct generation of mid-infrared radiation are highly attractive for portable broadband spectroscopic instrumentation in future terrestrial and space applications.**

The advent of optical frequency combs (OFCs) [1–5] has transformed the landscape of broadband molecular spectroscopy by surpassing the speed, resolution, brightness and frequency precision of traditional Fourier Transform Infra-Red (FTIR) spectrometers [6]. To date, high-resolution OFC systems based on interferometry [7,8], dispersive elements [9,10], and frequency filtering [11,12] have been demonstrated. Arguably, the most common detection scheme is the multiheterodyne approach of dual-comb spectroscopy (DCS), which has increasingly gained interest due to its attractive solid-state operation and fast acquisition speed [13]. In this scheme, two OFCs with different repetition rates, emitting light at equidistant and discrete optical frequencies, are heterodyned on a photodetector whose square-law characteristics produce a down-converted comb at radio frequencies (rf) [14,15]. Through this procedure, the frequency components of the optical spectrum are evenly dispersed over the electrical bandwidth of the photodetector rather than spatially dispersed or temporally separated as is the case for most other OFC detection schemes. This gives direct access to the entire spectrum with an unprecedented combination of short acquisition times and high signal-to-noise ratios.

Since a prerequisite for DCS is a pair of mutually coherent, broadband sources with narrow linewidths, in recent decades a surge of activity has been devoted to the development of novel comb generation schemes that have resulted in spectacularly broad coverages and fractional stabilities [4,16–18]. However, a vast majority of these require powerful external optical pumps to drive the nonlinear comb generation, and therefore very few contemporary OFC technologies are compatible with chip-scale integration and miniaturization for autonomous and portable DCS instruments. This challenge is further acuminated by the spectroscopic importance of the mid-infrared spectral region, where many molecular compounds have strong fundamental absorption features. Unfortunately, coherent sources and optical components at wavelengths beyond those of the telecommunication field are scarce, causing most mid-infrared DCS systems to rely on non-linear frequency conversion and parametric processes in bulk or periodically polled non-linear crystals (e.g. difference-frequency-generation based on periodically-poled lithium niobate (PPLN) crystals [19] or optical parametric oscillators (OPOs) [20,21]), which adds further complexity to the systems and limits their optical output power.

In this context, the chip-scale, electrically pumped quantum- and interband cascade laser (QCL/ICL) OFCs are particularly appealing as the basis for compact, solid-state, mid-infrared spectrometers that target applications favoring small footprint and high time-resolution over metrology-grade frequency accuracy and spectral coverage. Of the two, QCL-OFCs have reached a decidedly higher level of maturity, manifested by their recent commercial availability [22] and their inclusion in the first dual-comb spectrometer to reach the market [23]. However, for spectroscopic assessments in the 3-4 μm range where many hydrocarbons have characteristic absorption bands [15,19,17], the reduced efficiency of the QCL light generation mechanism, accompanied by difficult dispersion compensation [24,25], have thus far hampered comb formation.

This notable gap can be addressed by ICLs, which have shown room-temperature operation in the 3-6 μm range [26–29]. These devices are often viewed as hybrids of conventional diode lasers, since the gain is provided by interband transitions, and QCLs, due to the staircase architecture of the stages that lowers the threshold current density and parasitic voltage drop. This design makes the ICL an extremely energy-efficient mid-infrared source for spectroscopy in the exceedingly important C-H stretch region around 3.3 μm, which was recognized in NASA's selection of the ICL technology for methane detection in the Martian atmosphere during the Mars Curiosity mission [30]. Recent observations of seasonal variability in the methane concentrations measured by

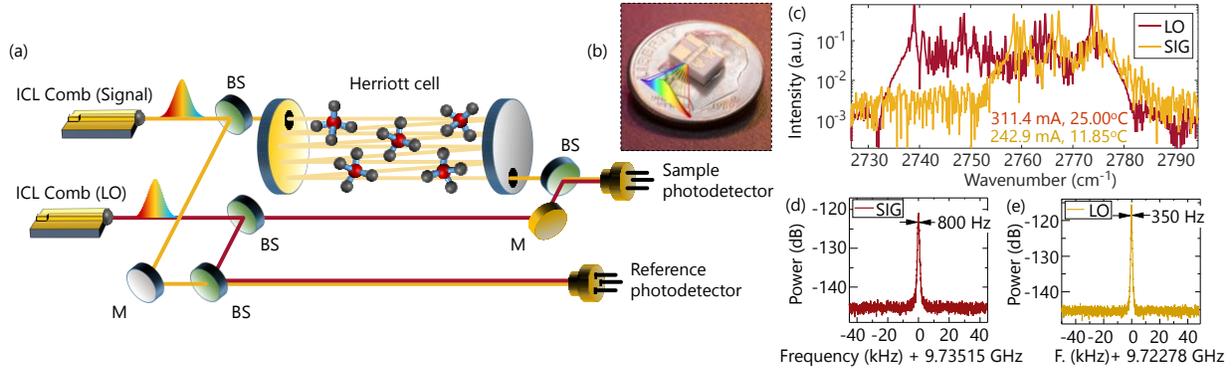

**Fig. 1.** Experimental setup and optical spectra. (a) Schematic for asymmetric (dispersive) dual-comb spectroscopy with a multipass cell. BS – beamsplitter, M – mirror. (b) An ICL-OFC on a beryllium mount. (c) Mode spectra in logarithmic scale for the two ICLs measured by an FTIR with 0.125 cm$^{-1}$ resolution. (d) Intermode beat note for the signal comb with 800 Hz 3 dB linewidth. (e) Intermode beat note for the local oscillator comb with 350 Hz 3 dB linewidth.

Curiosity have provided clues to the origin of Martian methane [31] and ways to merge the ICL's proven spectroscopic capabilities with broadband comb operation are now being pursued [32,33].

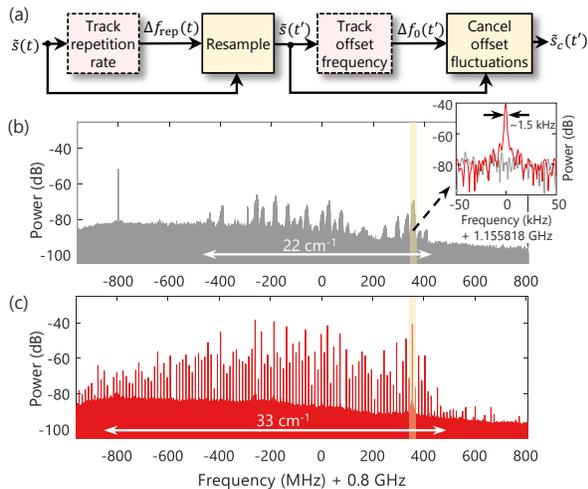

**Fig. 2.** Coherent averaging. (a) A simplified block diagram of the CoCoA algorithm. (b) rf multiheterodyne spectrum after 1 ms of acquisition time. (c) rf multiheterodyne spectrum after applying the CoCoA algorithm. The inset shows an expanded view of the rf beat note highlighted in yellow. (d) inset showing the linewidth of a typical beat note.

Here, we employ these advances in ICL comb technology in the first demonstration of an electrically-pumped dual-comb spectrometer with on-chip sources capable of high-sensitivity molecular detection in the important C-H stretch spectral region. The spectroscopic potential is experimentally verified through measurements of methane (CH$_4$) and hydrogen chloride (HCl) around 3.6 µm, over an optical bandwidth of ~33 cm$^{-1}$ (~1 THz) and with optical multipass (MPC) cell used for sensitivity enhancement enabling minimum detection limits (MDLs, 1σ) down to 0.1 ppmv/√Hz for HCl. The OFC mode spacing of 0.32 cm$^{-1}$ (9.7 GHz), which is dictated by the 4 mm laser cavity length, is well-suited for probing molecular structures containing C-H groups in the atmosphere. The two combs reported here require < 0.8 W of electrical drive power per device, which is far lower than any other mid-infrared OFC technology (including QCL-OFCs). This is promising for future low-power all-solid-state dual-comb spectrometers designed for terrestrial and space applications.

Figure 1(a) schematically illustrates the experimental configuration for the ICL-OFC dual-comb spectrometer. The ICLs were grown by molecular beam epitaxy on an n-GaSb substrate, using the design and growth procedures of ref. [33]. The ICL-OFCs are biased through 40 GHz rf probes (Cascade Microtech, ACP40-GSG-150) to enable extraction of the rf voltage characteristics through bias-tees (Sigatec, SB15D2). This exploits the intrinsic nonlinearities of the devices, which act as frequency mixers that produce distinct beat signals at the round-trip frequencies, the so called intermode beat notes [5]. Strong coherence between the modes is indicated by the narrow width of the intermode beat notes, which can be used to monitor the state of the comb operation during data acquisition. The lasers were operated at temperatures of 12°C and 25°C with bias currents of 311 mA and 242 mA, respectively, with bias voltages below 3 V. Figures 1(d) and (e) show the intermode beat notes extracted via the rf-ports of the bias-tees. Their sub-kHz linewidths indicate operation in the comb regime, and the center frequency difference of 12.4 MHz agrees well with the observed frequency spacing of the multiheterodyne spectrum shown in Fig. 2(c).

A high numerical aperture, AR-coated lens (Thorlabs 390037-E) was used to collimate approximately 8 mW of optical power from each device, and variable apertures were used to reduce optical feedback to the devices and to avoid detector saturation. The signal OFC was directed through a 76 m astigmatic Herriott MPC (Aerodyne Inc.), after which it was combined with the local oscillator (LO) OFC using a wedged CaF$_2$ beamsplitter (Thorlabs BSW510) and focused on a high bandwidth HgCdTe (MCT) photodetector (VIGO systems, 1 GHz). An additional matched-reference photodetector was also used to directly heterodyne the signal and LO OFCs. This provided intensity noise suppression through balanced detection, which has been shown to improve the performance of similar DCS systems based on QCL-OFCs [34,35]. To characterize the mode structure of the OFCs, the secondary beam from one of the beamsplitters was guided to an FTIR (Nicolet 8700) with 0.125 cm$^{-1}$ resolution. Figure 1(c) shows typical mode spectra for the two OFCs. The spectra indicate optical bandwidths of 45 cm$^{-1}$ and 33 cm$^{-1}$ for the LO and signal combs, respectively, with the spectral coverage of the dual-comb spectrometer being limited by the narrower comb. The outputs from the signal and reference photodetector were digitized using a 3.5 GHz oscilloscope with 8-bit vertical resolution (LeCroy

WavePro Zi735). The sampling frequency of 20 GS/s was down-sampled to 2.5 GS/s in post-processing, to increase the effective bit resolution and reduce the influence of digitization noise. Since the ICL-OFCs were operated in free-running mode, their poor frequency stability precluded measurements beyond the mutual coherence time of the lasers. This well-known issue in DCS can only be overcome by implementing coherent averaging schemes [13], which serve to align consecutive dual-comb interferograms either by active stabilization of the system or by computational techniques that do not require advanced stabilization hardware [36,37]. Here, we implement the latter by following the computational coherent averaging (CoCoA) scheme of ref. [37] [block diagram shown in Fig. 2(a)], in which the instantaneous frequency spacing of the rf comb is extracted by exploiting the self-mixing harmonics through digital difference frequency generation (DDFG). This information is used to re-sample the raw time-domain signals in post-processing, which ensures equal duration of the DCS interferograms. The second stage of the CoCoA procedure corrects for phase fluctuations between consecutive interferograms via a recursive frequency-tracking algorithm [38] that acts on the rf component with the highest SNR. Comparison between Fig. 2(b) and (c) shows the improvement in the multiheterodyne rf spectrum, which contains more than 100 beat notes with 3 dB linewidths of ~1.5 kHz [see Fig. 2(d)] and peak SNRs of up to 40 dB after correction. Clearly observable is the non-uniform amplitude distribution of the rf beat notes, which resembles the modulation caused by multiple solitons generated within a single roundtrip of a silicon microresonator [39]. This further justifies incorporation of the reference detector for the transmission measurements.

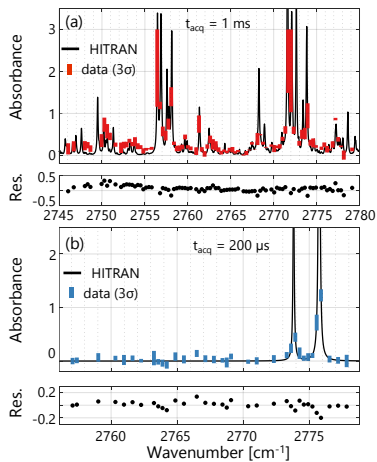

**Fig. 3.** Spectroscopy. (a) Absorbance spectrum of 0.4%$_v$ methane diluted in nitrogen. The measurement spans 33 cm$^{-1}$ with a frequency sampling interval of 0.32 cm$^{-1}$. The bottom panel shows the residual with a standard deviation of 12% (absorption units) for 1 ms of acquisition time. (b) Absorbance spectrum of 70 ppmv of hydrogen chloride diluted in nitrogen for 200 μs of acquisition time. The bottom panel shows the residual with a standard deviation of 9% (absorption units), which translates into a MDL (1σ) of ~0.1 ppmv/√Hz.

The spectroscopic detection sensitivity of the DCS system was evaluated through measurements of methane and hydrogen chloride diluted in nitrogen. For each measurement a known analyte concentration at atmospheric pressure was introduced into the MPC, and the photodetector signals from the signal and reference detectors were recorded. Due to limitations in the available acquisition memory, the duration of the measurements was limited to 1 ms. Figure 3(a) shows a measured absorbance spectrum of 0.4%$_v$ of methane at atmospheric pressure together with a model based on the HITRAN database [40]. Beat notes with SNRs below 8 dB have been excluded from the spectrum resulting in the optical bandwidth for this acquisition of 33 cm$^{-1}$ (1 THz). In general, good agreement with the HITRAN database is observed, although the 3σ error bars at these short acquisition time-scales are on the 10%-level (transmission units). The standard deviation of the residual is determined to be 12%, which translates into a MDL (1σ) for methane of 15 ppmv/√Hz. It should be noted that the absorption cross-sections of methane at 3.6 μm are relatively weak, with integrated linestrengths approximately 2 orders of magnitude weaker than those at 3.2 μm. Therefore, to investigate expected molecular detection limits for hydrocarbons when targeting their strong absorption bands, hydrogen chloride was also detected. For this measurement, 70 ppmv of HCl in N$_2$ was introduced to the MPC, which resulted in the absorbance spectrum shown in Fig. 3(b). The average SNR, based on the standard deviation of the baseline, is determined to be 770/√s, which equates to a MDL (1σ) of 0.1 ppmv/√Hz.

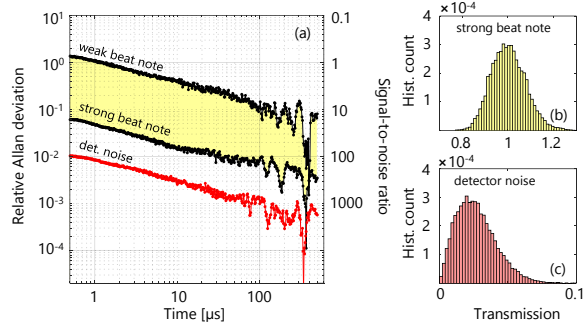

**Fig. 4.** Noise analysis. (a) Relative Allan deviation in transmission units as a function of acquisition time. The beat notes exhibit stabilities in the region highlighted in yellow. The red curve corresponds to the noise caused by the detector referenced to the strongest beat note. (b) Histogram of the acquired individual transmission values for the strong beat note. (c) Histogram of the transmission values for the detector noise. Note the asymmetry due to rectification of the detector noise when the histogram is recalculated into transmission units.

The short-term stability of the system was investigated by performing an Allan variance analysis, where the transmission values were extracted through computational IQ-demodulation with a 1 MHz bandwidth around the pertinent beat note. The results are shown in Fig. 4. The system exhibits white-noise-limited performance for acquisition times up to those allowed by the acquisition memory. The SNRs of the beat notes range from ~1,000/√s to ~10,000/√s for weak and strong beat notes, respectively. The detector noise level is a factor of 6 lower than that of the high SNR beat notes, where the additional noise is mainly the remaining intensity noise due to the finite common-mode rejection ratio (CMRR) of the balanced detection.

In summary, we have demonstrated a dual-comb spectrometer based on low drive power interband cascade laser frequency combs. The system incorporates an astigmatic Herriott multipass cell for chemical sensitivity enhancement and achieves minimum detection limits of less than 100 ppbv at one second integration time for hydrogen chloride, corresponding to a minimum absorption coefficient of <2×10$^{-7}$ cm$^{-1}$/√Hz. The interband cascade laser platform enables comb operation in the mid-infrared spectral region with an electrical power consumption of less than a watt, which is highly appealing for field applications that impose restrictions on the power and cooling capacity available to the system.

In the wake of the rapidly evolving fiber-coupled OFCs, whose spectral coverages and frequency resolutions will provide the basis for the most accurate and precise dual-comb spectrometers in the foreseeable future, a niche for miniaturized, fully-integrated, on-chip DCS systems capable of extended field use may emerge. The scalability of the interband cascade laser technology, together with its low electrical power consumption and demonstrated suitability for demanding applications [30,31], makes it one of the more promising candidates to fill this niche. Future advances in dispersion management accompanied by increases in the spectral coverage will serve to fully establish these lasers as suitable mid-infrared dual-comb spectroscopy sources for terrestrial and space applications.

**Funding.** DARPA SCOUT program (W31P4Q161001, and S804-14-034); Office of Naval Research (ONR), NASA JPL Graduate Fellowship Program, the Kosciuszko Foundation Research Grant.

**Acknowledgments.** We thank William Dix and Bert Harrop for their help with the laser fixtures and submounts.